# Design & Simulation of 128x Interpolator Filter

Rahul Sinha[1], Sonika[2]

[1] Dept. of Electronics & Telecommunication, CSIT, DURG, CG, INDIA
rsinha.vlsieng@gmail.com

[2] Dept. of Information Technology, CSIT, DURG, CG, INDIA,
sonika444@gmail.com

**Abstract.** This paper presents the design consideration and simulation of interpolator of OSR 128. The proposed structure uses the half band filers & Comb/Sinc filter. Experimental result shows that proposed interpolator achieves the design specification, and also has good noise rejection capabilities. The interpolator accepts the input at 44.1 kHz for applications like CD & DVD audio. The interpolation filter can be applied to the delta sigma DAC. The related work is done with the MATLAB & XILINX ISE simulators. The maximum operating frequency is achieved as 34.584 MHz.

## 1.    Introduction

The development of multimedia systems has increased the demand for an audio digital-to-analog converter (DAC) and the demand for low-cost, wide dynamic range and high linearity of a DAC. Because of its inherent benefit Delta-sigma modulation (DSM) is the most suitable DAC topology to satisfy these requirements. DSM can reduce the bits of input digital signal at the cost of increasing sample rate and a great many of digital circuits, but the area and the complexity of the analog part can be greatly reduced and the implement of digital circuit is very convenient by using developed digital signal process and CMOS technique. Therefore the cost of a DAC can be decreased and that's why the topology of DSM is popular now [2], [5].

A general block diagram for a sigma-delta D/A converter is depicted in Figure 1. It consists of four functionally different parts [1]. In the first phase, the sampling rate of the input discrete-time signal is increased using an interpolator filter. In the second phase, the noise shaper converts an n-bit input data stream into a 1-bit data stream. This data stream is converted in the third phase into an analog signal using a 1-bit D/A converter. The role of the noise sharper is to keep the noise generated by the quantization as low as possible in the baseband by shaping and moving this noise out of the baseband. The final phase consists of removing the out-of-band noise using an analog low pass filter.

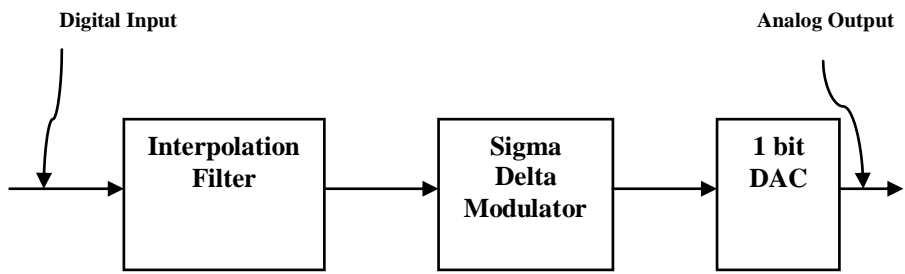

Figure 1 Block Diagram of Sigma Delta DAC

Figure 1 shows the system architecture of the audio DAC. It consists of three blocks, namely, the interpolator, the sigma-delta modulator and the 1-bit DAC. The audio DAC accepts PCM input data at sampling rates of 44.1 kHz. The interpolation ratio of the interpolator can be configured to 128x. For 44.1 kHz input signals, the interpolator gives the output data rate of 5.6448 MHz by setting the interpolation ratio as 128x. In this paper, we present an interpolator structure with its implementation in MATLAB. The Synthesis part is done for XILINX SPARTAN 6 chip.

The paper is structured as follows: Section 2 describes the basic principle of operation of Multistage Interpolation. The architecture of the implemented Interpolator is discussed in Section 3. Section 4 deals with experimental results of an implemented prototype are shown. Finally, we end up with some conclusions & future scope in Section 5.

## 2.    Interpolator Architecture

The function of interpolation filter is to raise the sampling frequency to oversampling rate (OSR * $f_s$) and to suppress the spectral replicas centered at $f_s$, $2f_s$, …, (OSR-1) $f_s$. Due to the high sampling rate, the pass-band and transition band are extremely narrow compared to the Nyquist bandwidth of the output signal, which means a single stage FIR filter to achieve 128x OSR has to be of exceedingly high order, so a multi-stage structure is preferred to reduce the computation complexity.

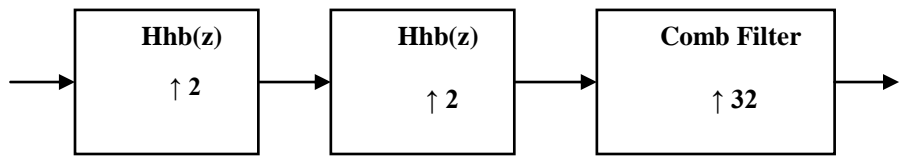

Figure 2 Interpolation Filter

The architecture of the 128x interpolation filter is shown in Figure 2. It is a multi-stage filter. The first two half-band filters increase the sampling rate of the signal by four times. The last stage is comb filter to provide 32 times sampling rate for input signals leading to overall interpolation ratio as 128x.

### 2.1.    Half Band Filter

Half band filters can be used for signal oversampling by 2 (x2). The resulting signal contains the original samples and the interpolated point between the original samples is the original samples filtered by the non-zero coefficients.

Linear-phase FIR half-band filters have found several applications in the past. For instance, in the design of sharp cutoff FIR filters, a multistage design based on half-band filters is very efficient. The efficiency of half-band filters derives from the fact that about 50 percent of the filter coefficients are zero, thus, cutting down the implementation cost. Half-cost filters have also been used in multirate filter bank applications, either directly or indirectly [4].

Let H(z) denote the transfer function of a (linear-phase, FIR) half-band filter of order N-1.

$$H(z) = \sum_{n=0}^{N-1} h(n)z^{-n}, \ h(n) real \qquad (1)$$

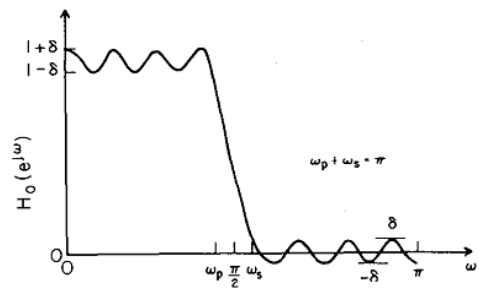

Figure 3 Typical amplitude Response of Half Band Filter

## 2.2. Comb/Sinc Filter

A comb filter provides the remaining factor of 32 increase in the sampling rate. The advantage of comb filter is their simple structure, which does not require any multiplier or coefficient storage, as compared with traditional FIR filters [3]. Cascaded integrator-comb (CIC) digital filters are computationally efficient implementations of narrowband lowpass filters and are often embedded in hardware implementations of decimation and interpolation in modern communications systems.

The comb filter has the transfer function of

$$T_{sinc}(z) = \frac{1}{M^N}\left(\frac{1-z^{-RM}}{1-z^{-1}}\right)^N, \qquad (2)$$

Where R = decimation or interpolation ratio, M= number of samples per stage, N = number of sections in filter.

The CIC filter's difference equation is:

$$y(n) = x(n) - x(n-D) + y(n-1), \qquad (3)$$

Usually a SincL+1 filter are used to filter out the quantization noise of an Lth order modulator.

A CIC interpolator would be N cascaded comb stages running at fs/R, followed by a zero-stuffer, followed by N cascaded integrator stages running at fs, where N represents the No. of sections (No. of Integrator Sections = No. of Integrator Section) [6].

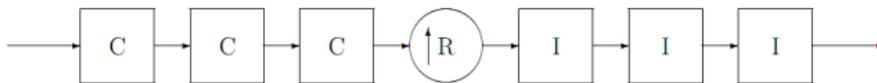

Figure 4 Comb Filter Structure having three sections

## 3.     Implementation

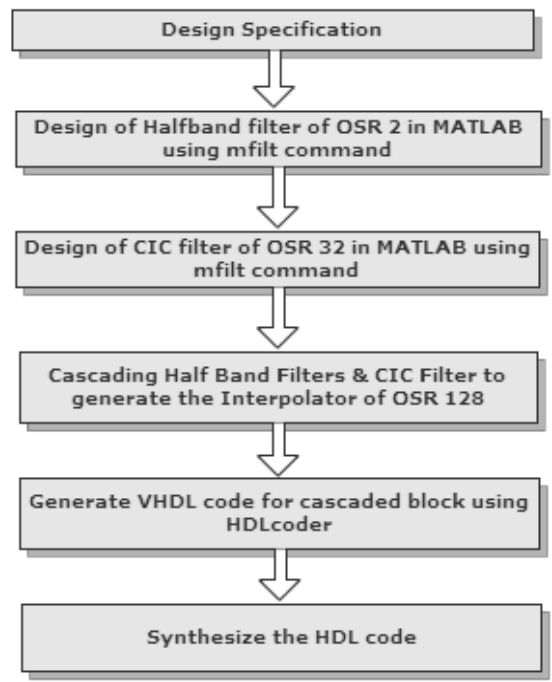

Figure 5 Flow Diagram of our work

In the implementation of the multistage interpolation we chose the specification as per Table I. Designing of the first & second stage Half Band filter is done with the mfilt of MATLAB. Similarly the CIC filter of the OSR 32 is achieved with the mfilt command. The three filters is then cascaded and the relevant VHDL code is generated with the generateHDL (HDL Coder) in MATLAB.

The VHDL code is exported to the Xilinx ISE 13.1 where the code is synthesized to get the hardware requirement for the proposed 128x interpolator.

Table I Specification of Sub Blocks

| Stage | Type of filter | OSR | Sampling Frequency / kHz | Stop-band attenuation / dB |
|---|---|---|---|---|
| 1 | Half-band | 2 | 44.1 | 80 |
| 2 | Half-band | 2 | 88.2 | 80 |
| 3 | Comb/ Sinc | 32 | 5,644.8 | 65 |

## 4.     Experimental Results

The complete design has been prepared with MATLAB & Xilinx ISE 13.1. The Coefficients generation of each filter is generated using the MATLAB and then the VHDL code is generated with the help of HDL Coder. Generated VHDL code is synthesized for the XILINX SPARTAN6.

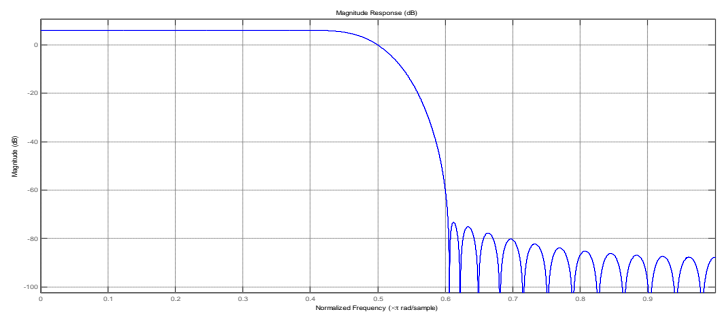

Figure 6 Magnitude Response of Halfband Filter

Figure 6 shows the Magnitude response for the first & second stage of the Interpolator. Figure 7 depicts the Signals at input of the HBF1 & at the output of HBF1 & HB2.

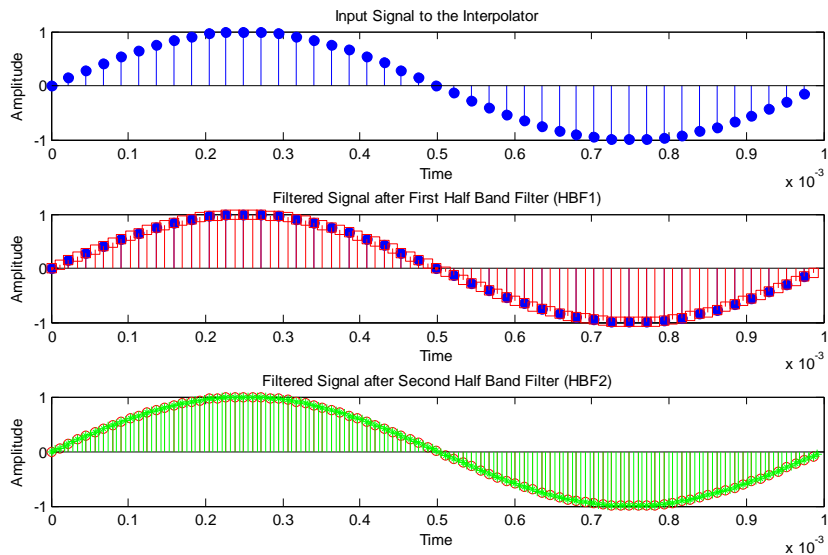

figure 7 Signals at the input of Interpolator & at the output of HB1 & HB2

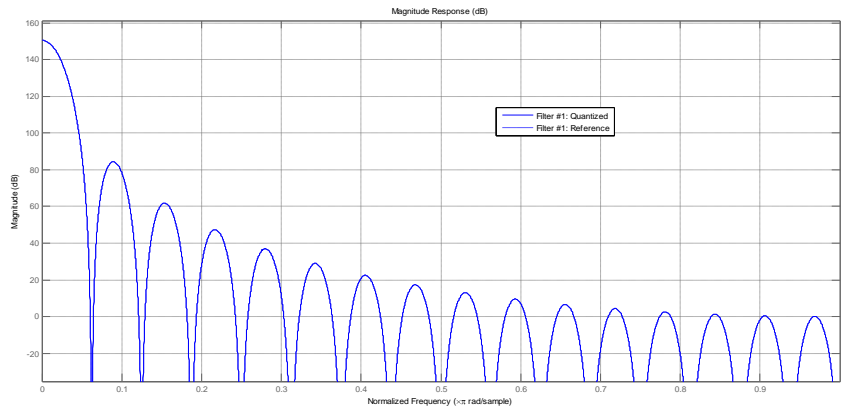

Figure 8 Magnitude Response of Comb/Sinc Filter

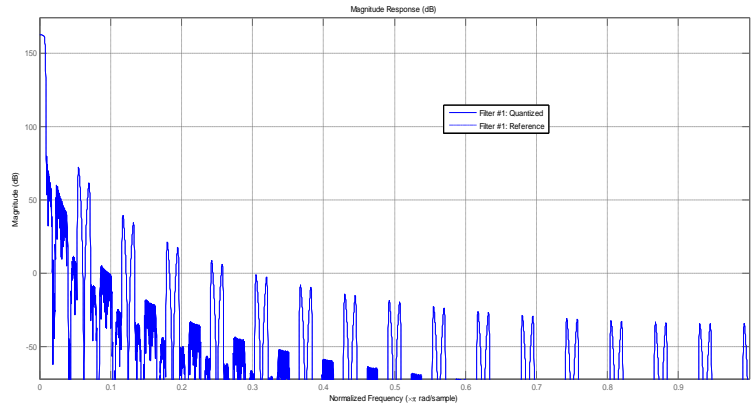

Figure 9 Magnitude Response of Interpolator

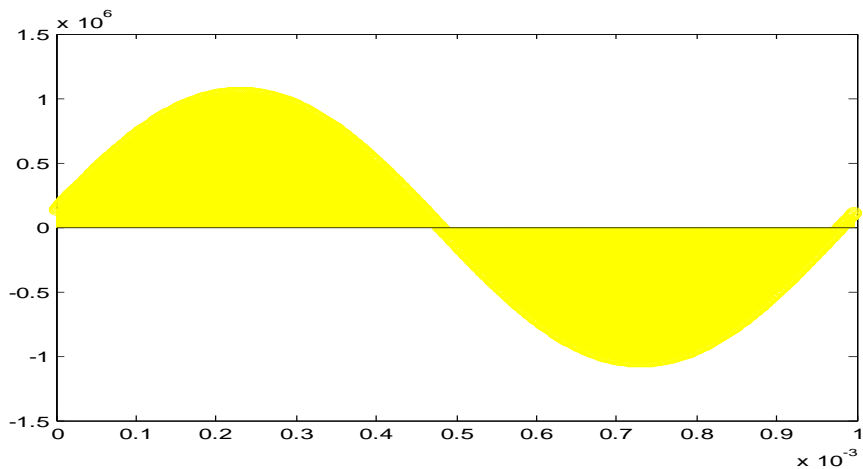

Figure 10 Signals at the output of Interpolator

The designed interpolator is implemented for XILINX VERTEX 7, device XC7VX330T, package FFG1157 with speed grade -3. Table II shows the implementation details of our design. Each row of the table shows the hardware cost of each stage of Interpolator as per specification mentioned in Table 1.

The last row shows the hardware implementation cost of the interpolator. The maximum achievable frequency of operation for the interpolator is 34.584 MHz.

Table II Synthesis Result of Blocks for XILINX VERTEX 7

| Design | Cascade Design of HBF1, HBF2 & Sinc Filter |
|---|---|
| **No. of Slices Register** | 1037 out of 408000 |
| **No. of Slice LUTs** | 384 out of 204000 |
| **No. of Bonded IOBs** | 40 out of 600 |
| **Max. Clock Frequency** | 34.584 MHz |
| **No. of Multiplier** | 48 |
| **No. of Adder** | 61 |
| **No. of Register** | 63 |
| **No. of Multiplexer** | 8 |

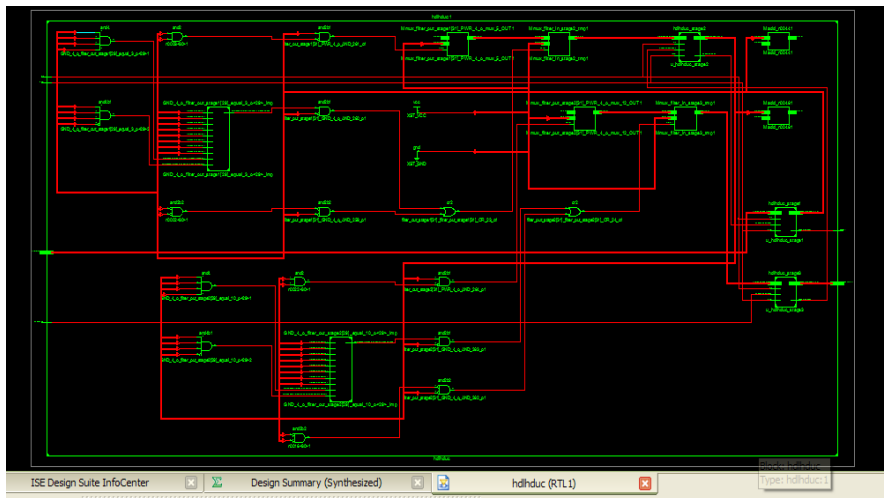

Figure 11 RTL view of Interpolator (XILINX VIRTEX 7)

## 5.   Conclusion & Future Scope

A method to construct the interpolation filter of Sigma Delta Audio DAC is presented in this paper. Cascade halfband filter and a Comb/Sinc filter comprise the interpolation filter. Multirate filtering & HDL Coder feature of the MATLAB is utilized to achieve the design.

The optimization can be performed so that the overall interpolator contains no general multipliers. This is achieved by using hardware efficient FIR filters in a tapped cascaded interconnection of identical sub-filters, which requires no multipliers. Further area optimization is achieved by the multiplier less CSD encoding.

## References

1. Cheung Ray, Pun KP, Yuen SCL, Tsoi KH, Leong PHW, "An FPGA based Re-configurable 24-bit 96kHz Sigma-Delta Audio DAC", Hong Kong, 2003
2. Li J., Wu X.B., Zhao J.C., "An Improved Area Efficient Design Method for Interpolation Filter of Sigma-Delta Audio DAC", IEEE, 2010
3. Yunfeng P., Kong D., Feng Z., "Design and Implementation of a Novel Area Efficient Interpolator", Chinese Journal of Semiconductors, Vol. 27 No. 7, CHINA, July 2006
4. Vaidyanathan P.P., Nguyen T.Q., "A Trick for the Design of FIR Half Band Filters", IEEE Transcations on circuits and systems, Vol. CAS -34, No. 3, March 1987
5. Yang W.R., Cheng Y.Y., Wang J.M., "Simulation of Multi bit Digital Delta Sigma Modulator", ICEPT-HDP, IEEE, CHINA, 2008
6. Donadio M. P., "CIC Filter Introduction", Iowegian, July 2000
7. Mathworks.com
8. XILINX.com
9. Pedroni V.A., "Circuit Design with VHDL", MIT Press, Cambridge, London, England, 2004